\documentclass{eph}

\title{Universal Mechanisms in the Growth of Voluntary\\ Organizations}
\shorttitle{Growth of Voluntary Organizations}
\author{Fredrik Liljeros\inst{1,2} \and Lu\'{\i}s A. Nunes
Amaral\inst{3} \and H. Eugene Stanley\inst{4}}
\shortauthor{Liljeros, Amaral, and Stanley}
\institute{
  \inst{1} Department of Sociology, Stockholm University,
         S-106 91 Stockholm, Sweden \\ 
  \inst{2} Swedish Institute for Infectious Disease Control,
         SE-171 82 Solna, Sweden\\
  \inst{3} Dept. of Chemical Engineering, Northwestern University,
Evanston, IL 60208, USA \\
  \inst{4} Center for Polymer Studies \& Dept.~of Physics,
  Boston University, Boston, MA 02215, USA
}
\usepackage{pxfonts}


\begin{document}

\maketitle

\begin{abstract}
We analyze the growth statistics of Swedish trade unions and find a
universal functional form for the probability distribution of growth
rates of union size, and a power law dependence of the standard
deviation of this distribution on the number of members of the union.
We also find that the typical size and the typical number of local
chapters scales as a power law of the union size.  Intriguingly, our
results are similar to results reported for other human organizations of
a quite different nature.  Our findings are consistent with the
possibility that universal mechanisms may exist governing the growth
patterns of human organizations.

\end{abstract}

\section{Introduction}

The developments of the last decade and a half in the former Eastern Block
countries show that democratization and market reforms do not automatically
generate healthy democracies and healthy market economies.  Several studies
suggest that the reason for this failure may be that well-functioning
societies \cite{Putnam94}
are fostered by the existence of a dense web of voluntary
organizations
which facilitate the creation of a ``social capital of trust'' among the
members of the society.  Indeed, many studies support the importance of
institutional settings for the maintenance of healthy societies. A telling
example is a study of the functioning of democratic institutions in the 27
regions of Italy, suggesting a correlation between a dense web of small
voluntary organizations and a dynamic civil society \cite{Putnam94}.
Specifically, in regions where people are embedded in a rich environment of
decentralized civic networks, there is an increasing likelihood that the
individuals will be able to cooperate in endeavors of mutual benefit
\cite{DethPutnam}. It would appear then that voluntary organizations promote
the creation of a ``social capital of trust'' that helps serve the functions
of (a) overcoming the anonymity of life in large societies, which may breed
``free-rider'' behavior \cite{Putnam00a}, and (b) overcoming the difficulties
in partitioning the exploitation of public resource \cite{Hardin}, such as
use of public water resources, limitation of air emissions, or determination
of fishing quotas.

Because of their societal importance, research on voluntary organizations has
been very active, including many different aspects such as (i) competition
between voluntary organizations and other organizations
\cite{McPhersonHannan}, (ii) the impact of social networks in membership
recruitment \cite{McAdam96}, (iii) social background of members
\cite{Mcpherson81}, and (iv) the ability of voluntary organizations to
overcome ``free-rider'' behavior \cite{Mancur65}. Despite this research
activity, one area that has not been pursued concerns the quantitative
characterization of the growth dynamics of voluntary organizations.

%
\begin{figure}[b]
 \vspace*{0.cm}
 \includegraphics*[width=6.5cm]{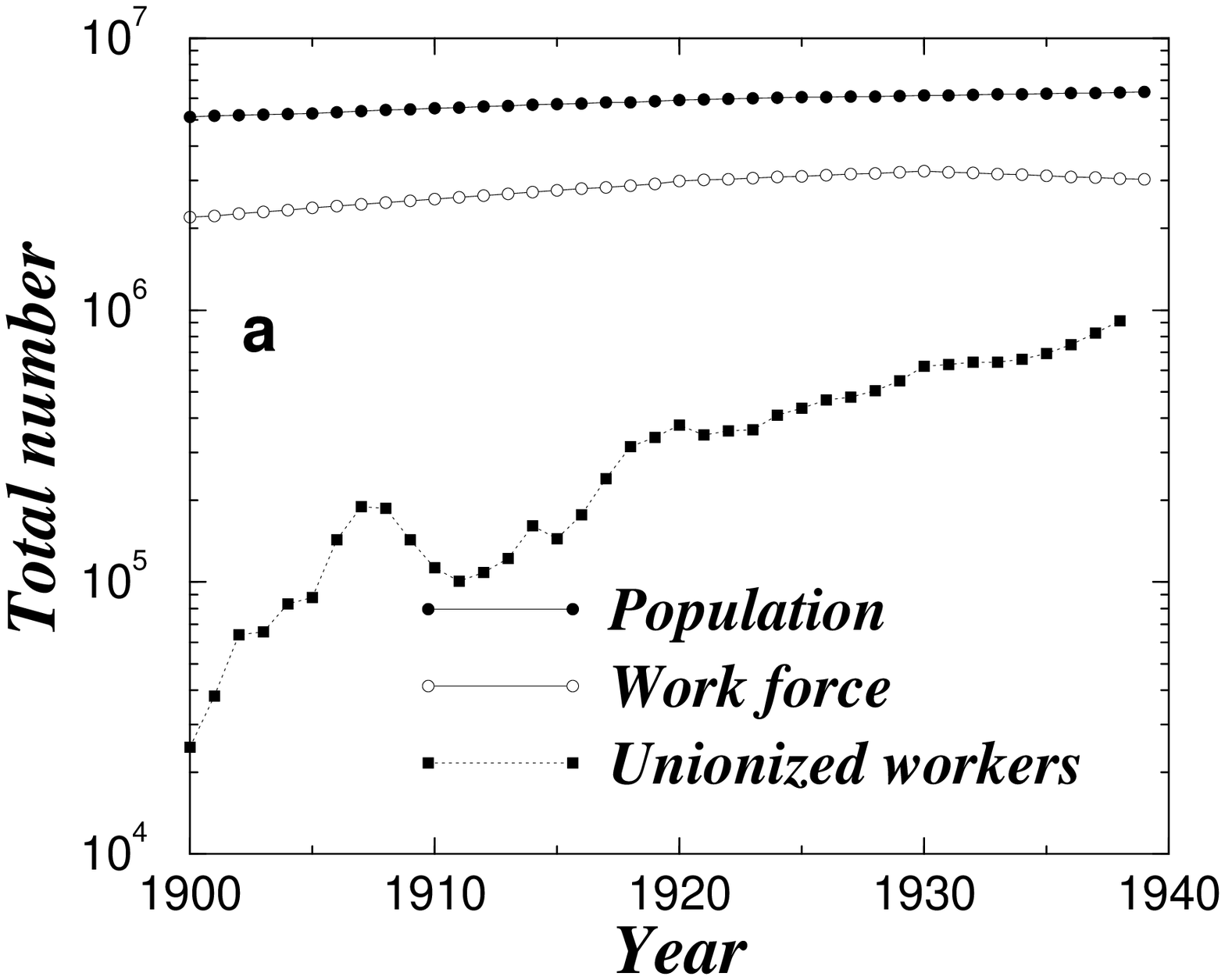}
 \includegraphics*[width=6.5cm]{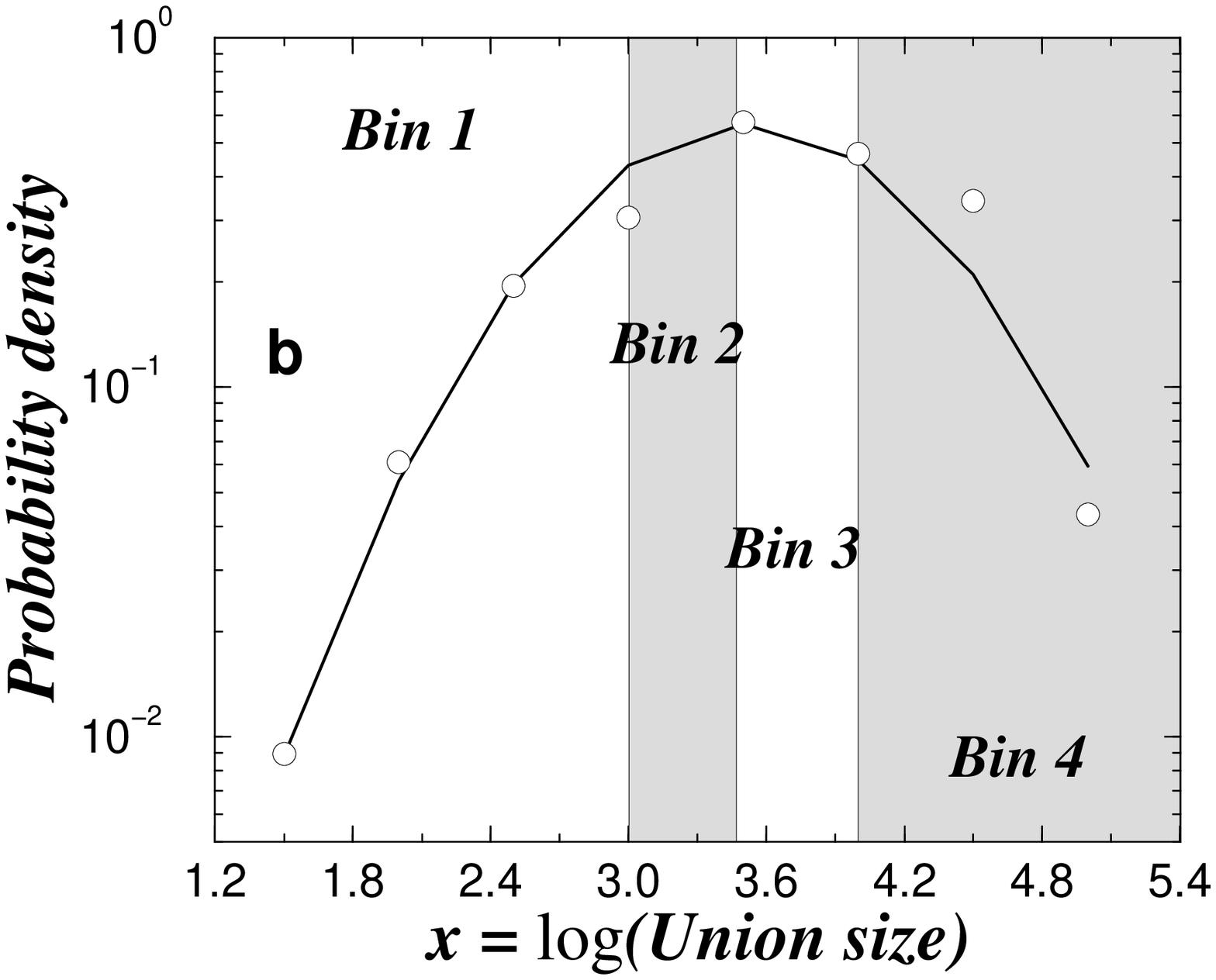}
 \vspace*{-0.cm}
 \caption{Historical data for total number of union members in Sweden.  (a)
  Time evolution of Swedish population, work force, and unionized work force
  for the period 1900--1940. In the subsequent analysis, we deflate the
  number of members of a union---its size---by the population growth, to
  remove the effect of population growth on the analysis. 
  (b) Probability density function of the size Swedish trade unions for all
  years and all unions. The distribution can be well approximated by a
  log-normal fit (full line in figure), which suggests, according to Gibrat's
  theory \protect\cite{Sutton97}, that the growth is a random multiplicative
  process. For subsequent analysis, we partition the data into 4 bins
  according to union size, as illustrated by the figure. }
 \label{f-background}
\end{figure}
%

Here, we use concepts and methods of statistical physics
\cite{Vicsek92,Stanley99} to quantify the growth of voluntary
organizations. Specifically, we test the possibility (i) that the statistical
properties of fluctuations in the output of a system yield information
regarding the underlying processes responsible for the observed macroscopic
behavior \cite{Vicsek92,Stanley99}, and (ii) that the precise details of the
interaction between the subunits comprising the system may play virtually no
role in determining the macroscopic behavior of the system
\cite{Stanley99}. A striking example is the behavior of response functions in
the vicinity of the liquid-gas critical point (the temperature and density at
which liquid and gas become indistinguishable fluids)
\cite{Vicsek92,Stanley99}.  Close to their respective critical points, very
different liquids---such as water, a polar molecule that forms hydrogen
bonds, and argon, an inert atom---become extremely sensitive to disturbances
yet their responses to those disturbances have identical spatial and temporal
scale-invariant properties.


%
\begin{figure}
 \vspace*{0.cm}
 \includegraphics*[width=6.5cm]{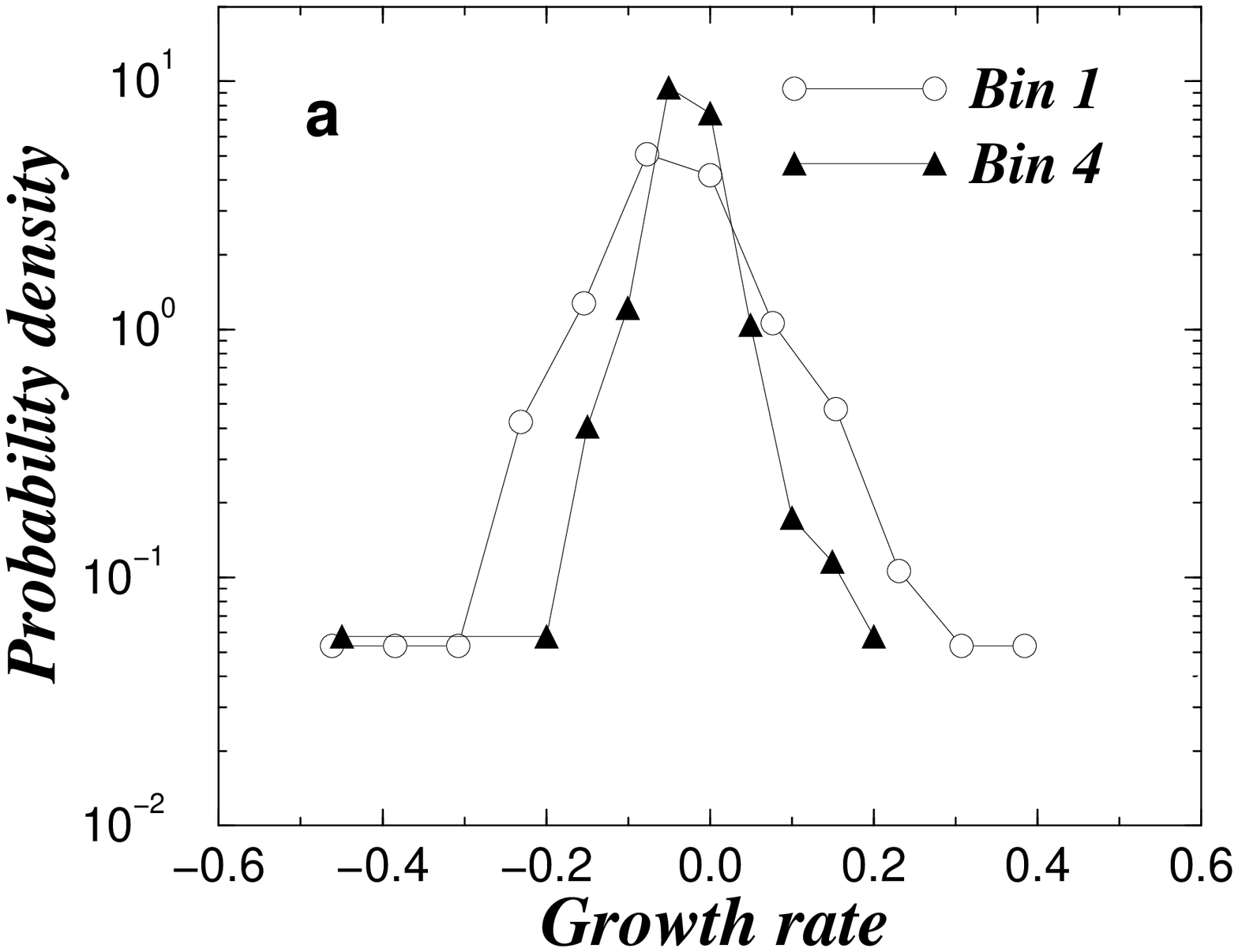} 
 \includegraphics*[width=6.5cm]{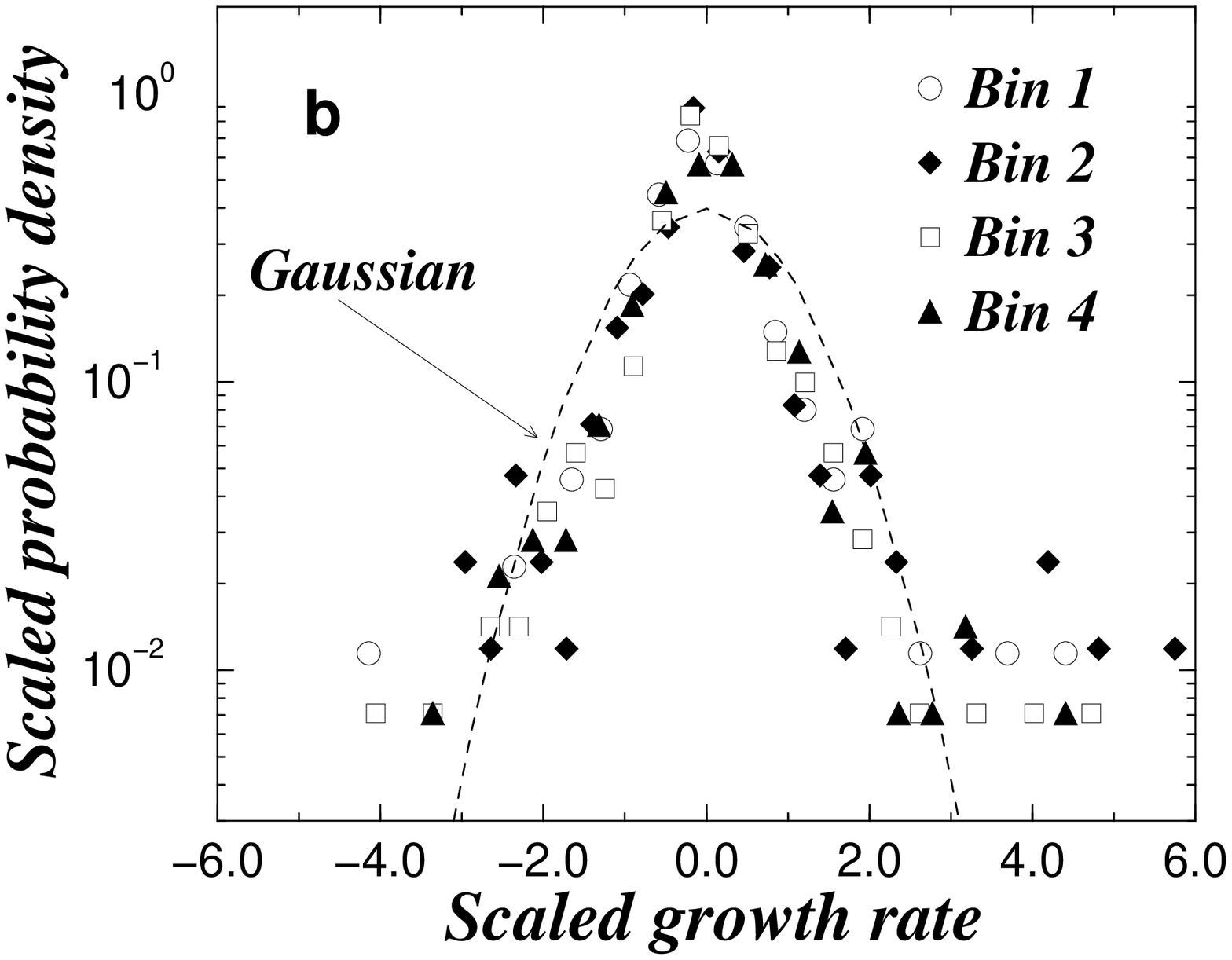}
 \vspace*{0.0cm}
 \caption{ (a) Probability density function for two different size bins. The
  annual growth rates of the unions are ``normalized'' by subtracting the
  average growth rate for all trade unions in the specific year. For clarity,
  we plot only the results for bins 1 and 4. The figure reveals two
  interesting points. First, the width of the distribution of growth rates
  decreases with the size of the union---e.g., the width is smaller for bin 4
  which comprises the largest unions. Second, the distribution of growth
  rates does not appear to be Gaussian. This second result is intriguing
  because each union comprises several local chapters (or sections), so one
  might expect that the central limit theorem applies---leading to a Gaussian
  distribution of growth rates.
  (b) The non-Gaussian character of the distributions is clearer in this
  log-linear plot showing the distributions of union growth rates for all 4
  bins scaled by the standard deviations of the corresponding
  distributions. The figure also suggests that the functional form of the
  distribution is {\it independent\/} of the size of the union. }
  \label{f-grates}
\end{figure}
%

\section{Results}

We analyze a  database \cite{Source} that provides a detailed
resource for the study of the growth statistics of a range of different
Swedish voluntary organizations---including trade unions
\cite{Hedstrom94}, temperance movements \cite{Sandell98}, free churches
\cite{Stern99}, and the social democratic party
\cite{Hedstrom00}---during the 50 yr period 1890--1940. We concentrate
our study on trade unions for three reasons. First, there are over
10,000 local chapters (or sections) comprising 60 trade unions, while
there are only 5 free churches, 5 temperance movements and 1 political
party in the database. The larger number of trade unions enables us to
make a more significant statistical analysis of the growth process for
the organizations. Second, a number of studies \cite{Rothstein00}
indicate that Swedish trade unions played a very important role in the
democratization process in Sweden, making their study particularly
relevant. Third, unions are a particularly interesting type of
voluntary organization as the decision to join a union is not an easy
one: A prospective new member would ideally balance (i) the benefit of
avoiding social pressure from fellow workers and (ii) hoped-for
long-term benefits of membership, against (i) the investment of time and
money into the organization, and (ii) the risk of losing the job or of
being discriminated against.


We start by defining the annual growth rate---that is to say, the size
fluctuation---of a union,
\begin{equation}
g(t) \equiv \log \left( \frac{S(t+1)}{S(t)} \right) \,,
\end{equation}
where $S(t)$ and $S(t+1)$ are the number of members of a given union in the
years $t$ and $t+1$, respectively, deflated by Sweden's total population. We
find that the statistical properties of the growth rate $g$ depend on $S$;
the magnitude of the fluctuations $g$ will decrease with $S$ since large
organizations have smaller relative fluctuations. We partition the trade
unions into bins according to their number of members---the union size
(Fig.~\ref{f-background}).  Figure~\ref{f-grates}(a) suggests that the {\it
conditional\/} probability density, $p(g | S)$, has the same functional form,
with different widths, for all $S$. To test whether $p(g | S)$ has a
functional form independent of union size, we plot the scaled quantities:
$\sigma(S) p(g/\sigma(S) | S)$ versus $g/\sigma(S)$. Figure~\ref{f-grates}(b)
shows that the scaled distributions ``collapse'' onto a single curve
\cite{Vicsek92,Stanley99}, consistent with the functional form
\begin{equation}
p(g | S) \sim \frac{1}{\sigma(S)} ~ F \left( \frac{g}{\sigma(S)} \right)\,.
\end{equation}

\begin{figure}
 \vspace*{0.cm}
 \includegraphics*[width=6.5cm]{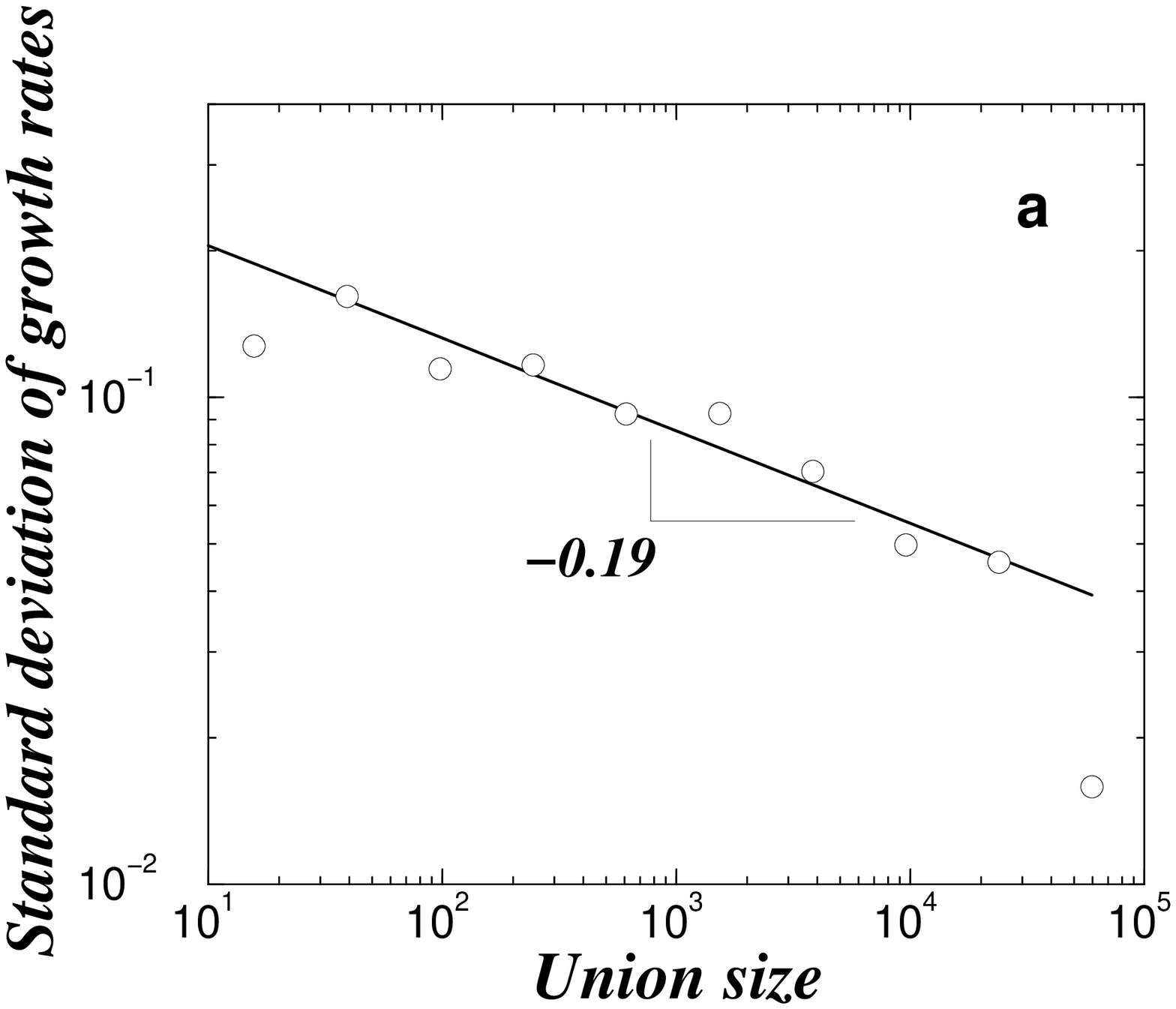}
 \includegraphics*[width=6.5cm]{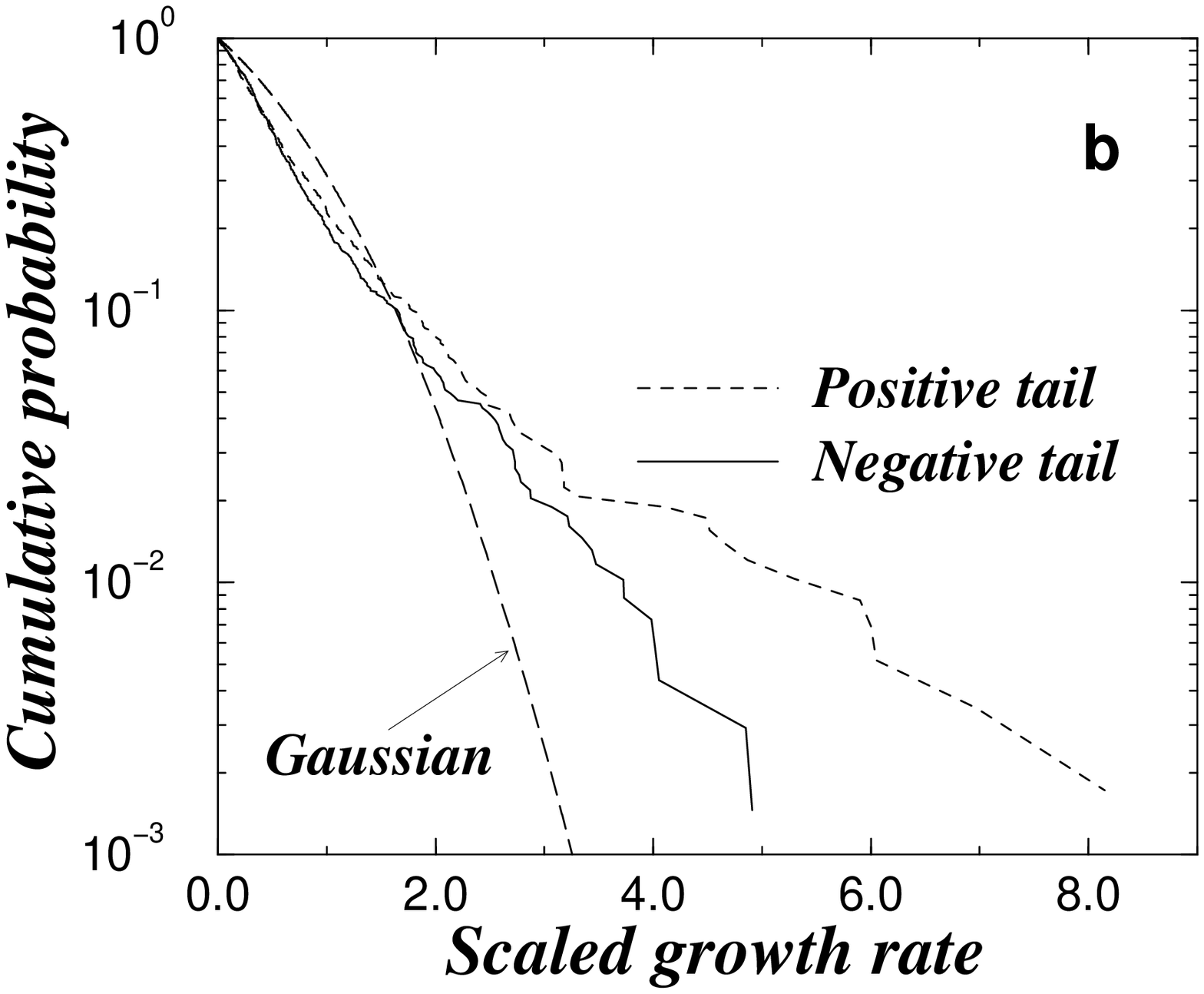} 
 \vspace*{-0.cm}
 \caption{ (a) Dependence of the standard deviation of the distribution of
  growth rates on number of members of the union. The fact that power law
  dependence of the standard deviation on size holds over three orders of
  magnitude---from unions with 40 members to unions with 40,000
  members---suggests that this finding is not spurious. The straight line
  is a power law fit to the region $40 \le S \le 40,000$ yielding an exponent
  estimate $\beta = 0.19 \pm 0.05$.
  (b) Functional form of the distribution of growth rates. We plot the
  cumulative distribution of the scaled growth rates from all bins. The
  cumulative distribution, which yields the probability of finding values
  larger than a certain threshold, is obtained by integrating the probability
  distribution function between the threshold value and infinity. The figure
  confirms that the distribution is not Gaussian, but may be consistent with
  either an exponential or stretched exponential dependence in the tails.}
 \label{f-grates1}
\end{figure}

We next calculate the standard deviation $\sigma(S)$ of the distribution
of growth rates as a function of $S$. Figure~\ref{f-grates1}(a)
demonstrates that $\sigma(S)$ decays as a power law
\begin{equation}
\sigma(S) \sim S^{-\beta},
\label{e-sigma}
\end{equation}
with $\beta = 0.19 \pm 0.05$. 

%
%


We next address the question of how to interpret our empirical results.  We
first note that a union is comprised of several local chapters, spread around
the country. A reasonable zeroth-order approximation is that the number of
members of the different local chapters comprising a given trade union will
grow independently; so the growth of the size of each union as the sum of the
independent growth of local chapters with different sizes. In a
recently-proposed model \cite{Amaral98,Sutton97}, the subunits comprising the
organization grow by an independent, Gaussian-distributed, random
multiplicative process with variance $v^2$. Existing subunits are absorbed
when they become smaller than a ``minimum size'', which is a function of the
activity they perform. Subunits can split into two new subunits if they grow
by more than the minimum size for a new subunit to form. The model predicts
$\beta = v / [2(v + w)]$, where $w$ is the width of the distribution of
minimum sizes \cite{Amaral98}.


\section{Internal organization of the unions}

\begin{figure}
 \vspace*{0.cm}
 \includegraphics*[width=6.5cm]{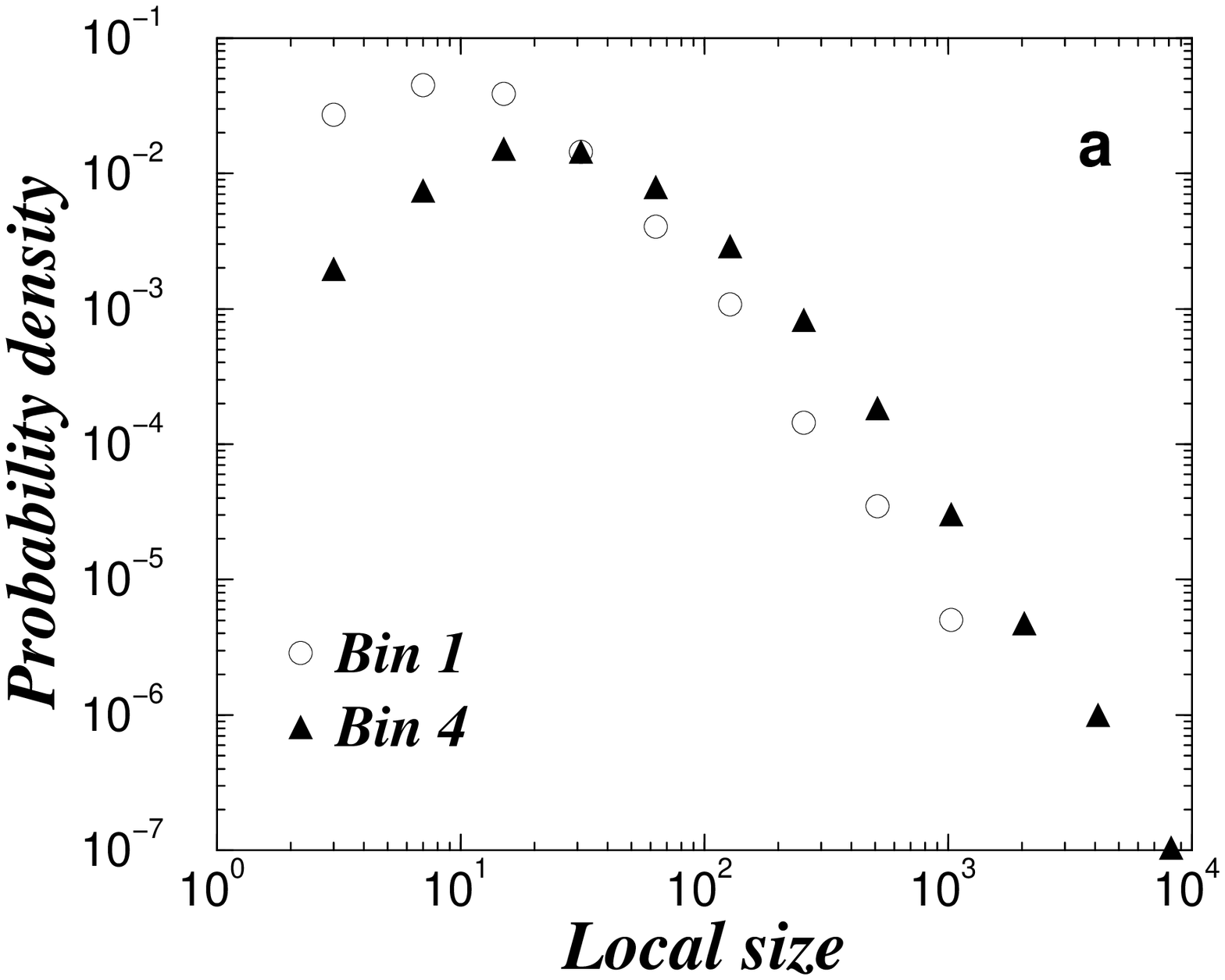}
 \includegraphics*[width=6.5cm]{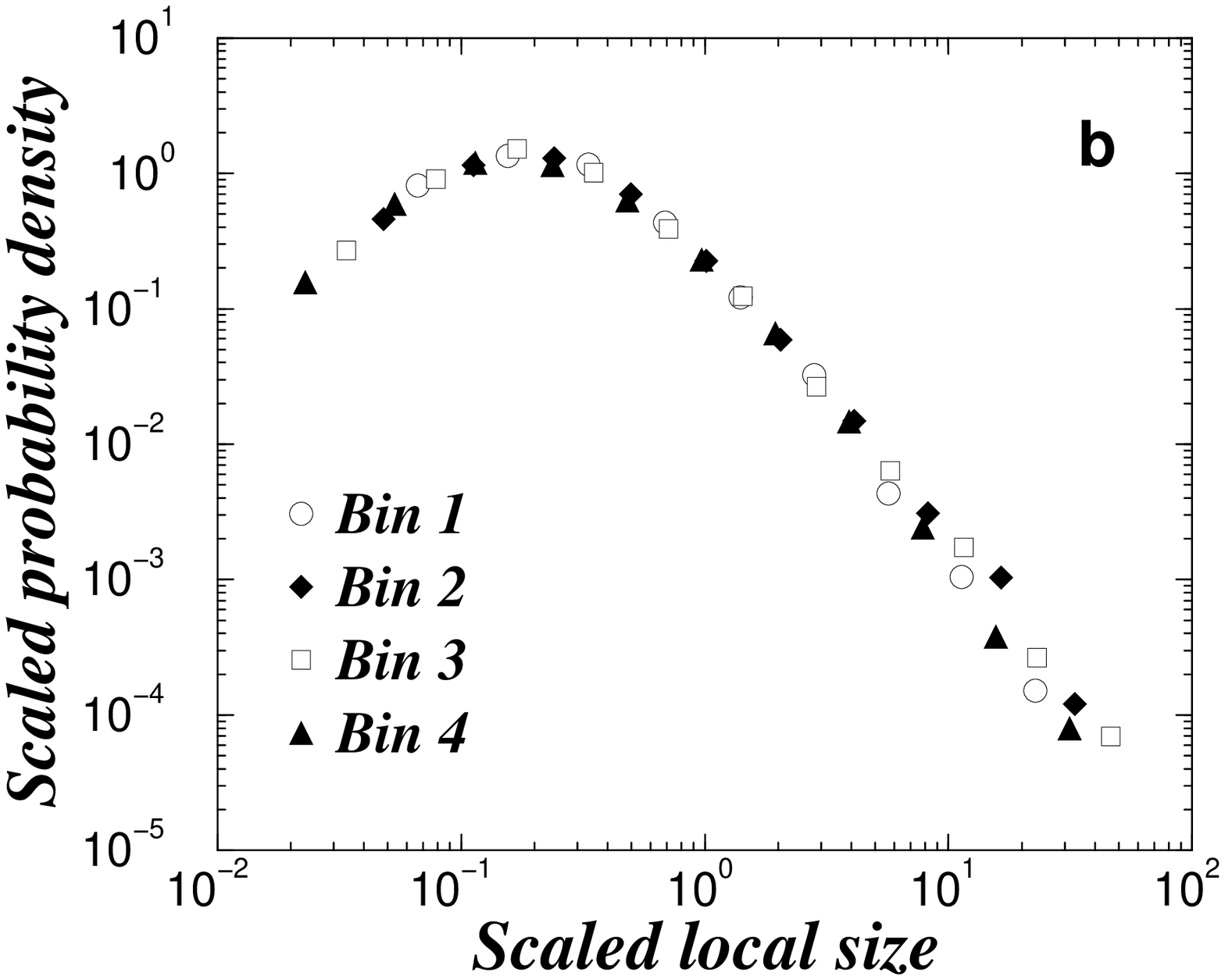}
 \vspace*{-0.cm}
 \caption{ (a) Probability density function of number of members of a local
  chapter, conditional on the size of the union it belongs to. We plot our
  results for two bins of union size. The figure reveals two interesting
  points: (i) the typical size of the local chapters increases with union
  size, (ii) the functional form of the distribution appears to be
  independent of union membership.
  (b) Scaled probability density function of scaled local chapter size
  conditional on union size (see text immediately following
  Eq.~(\protect\ref{e-self}) for details). The data for the four bins
  collapse onto a single universal curve, suggesting that the the structure
  of different unions is independent of union size except for a scale
  factor. }
 \label{f-internal}
\end{figure}

\begin{figure}
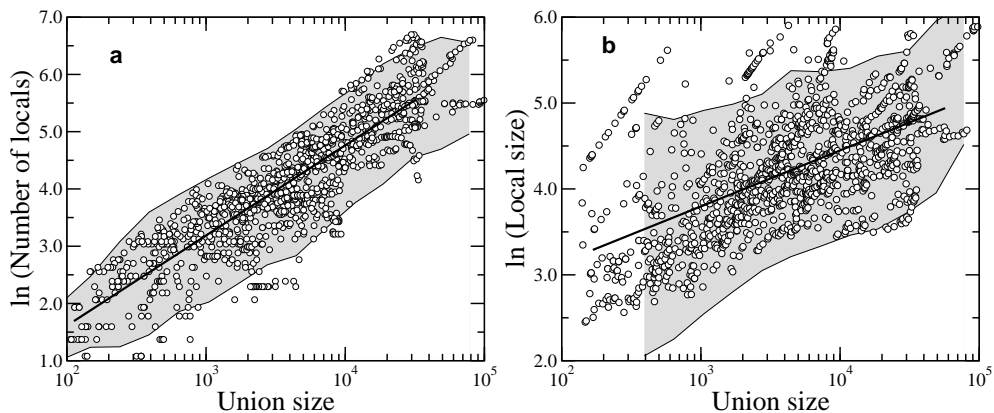

 \vspace*{0.cm}
 \includegraphics*[width=6.5cm]{liljeros2-fig-nlocals.eps}
 \includegraphics*[width=6.5cm]{liljeros2-fig-size-locals.eps}
 \vspace*{0.cm}
 \caption{ (a) Number $n$ of local union chapters comprising a given
  trade union as a function of union size.  The thick continuous line
  is a linear least squares fit between $\ln n$ and $\ln S$.  The gray
  areas defines the 95\% confidence interval. The data surprisingly
  fall along a straight line in a log-log plot, indicating a simple
  power law dependence on the union size of the number of local
  chapters comprising a union, with an exponent $\alpha = 1 - 0.68 \pm
  0.05 = 0.32 \pm 0.05$.
  (b) Size of local union chapters comprising a given trade union as a
  function of union size. The thick continuous line is a linear least
  squares fit between $\ln\xi$ and $\ln S$.  The gray areas defines
  the 95\% confidence interval.  We find a similar power law dependence
  of the typical size of the local chapters comprising a union with a
  given size with an exponent $\alpha = 0.3 \pm 0.05$. Note that the
  two independent estimates of $\alpha$ are within error bars. }
 \label{f-internal1}
\end{figure}

As trade unions have a complex internal structure, it is natural to
inquire what are the statistical properties of the set of local
chapters comprising a given union \cite{Amaral98,Plerou99,Keitt02}. To
this end, we quantify how the internal structure of a trade union
depends on its size. Specifically, we calculate the conditional
probability density $\rho(\xi|S)$ to find a local chapter with $\xi$
members in a union with $S$ members; Fig.~\ref{f-internal}(a). The
model predicts that $\rho(\xi|S)$ obeys the scaling form
\cite{Amaral98}
\begin{equation}
\rho(\xi|S) \sim \frac{1}{\xi_t(S)} ~ F_1 \left( \frac{\xi}{\xi_t(S)}
\right) \,,
\label{e-self}
\end{equation}
where $\xi_t(S) \sim S^{\alpha}$ is the typical size of a local chapter
in a trade union of size $S$, and $F_1(u)$ appears to decay as a stretched
exponential or a power law for $u \gg 1$. As a test of the scaling
hypothesis (\ref{e-self}), we plot the scaled quantities $S^{\alpha}
\rho(\xi|S)$ versus $\xi / S^{\alpha}$ and obtain a good data collapse,
that is, all data points fall onto a single universal curve;
Fig.~\ref{f-internal}(b). To estimate $\alpha$, we use the fact that
Eq.~(\ref{e-self}) implies that the typical {\it number\/} of local
chapters in a trade union with $S$ members increases proportionally to
$S^{1-\alpha}$ with $\alpha = 0.32 \pm 0.05$, while the typical number
of members of these local chapters is proportional to $S^{\alpha}$ with
the independent estimate $\alpha = 0.30 \pm 0.05$;
Figs.~\ref{f-internal1}(a),(b).

%
%

%
%


\section{Discussion}

Which characteristics of a voluntary organization are important for the
creation of social capital is a subject of debate in the literature
\cite{Levi96}. One may reasonably hypothesize that the size of the
subunits will be {\it negatively\/} correlated with their capacity for
creating social capital, since members in large subunits will likely (i)
not be able to create strong links among one another, and (ii) not be
able to participate in the governing process as fully as the members of
small subunits.  Our results support this hypothesis and suggest that
large organizations---because they typically consist of larger
subunits---will be less effective in creating social capital than small
organizations. This result may find support in the current trend in
high-tech firms to organize projects around small teams that split, when
they become too large, in order to facilitate cooperation.

Our findings are also of note for other reasons: First, our approach
differs from the statistical methods traditionally used in
macrosociology, which typically assumes that systems are linear and in
an equilibrium state \cite{Sorensen98}. It also stands in contrast to
the view that sociological explanations ideally would only make reference to
individual agents and their actions (``methodological individualism''
\cite{Putnam94}). We show that techniques successfully used in
statistical physics can be applied to a central sociological
topic---voluntary organizations---to reveal nontrivial patterns and
relationships.

Second, an intriguing aspect of our findings is that they provide
evidence for growth dynamics similar to those found for other
organizations, such as business firms \cite{Stanley96}. This similarity
is rather surprising as the reasons for the growth of a voluntary
organization are quite different from those for a business firm. In
particular, the profit motive---perhaps the most important factor in the
growth of business firms---is not evident for voluntary
organizations. The similarity between the empirical laws describing the
growth of voluntary organizations and business firms \cite{Stanley96},
and the fact these two types of organizations are apparently so
different, raises an intriguing analogy between the growth of human
organizations comprised of many {\it animate\/} interacting units and
the physics of natural systems comprised of many {\it inanimate\/}
interacting units. Our findings are consistent with the possibility that
universal mechanisms governing the growth of human organizations---such
as the complex internal structure of units, stochastic growth, and a
broad range of scales---are more important than the idiosyncratic
characteristics of the system that are customarily believed to determine
the system's dynamics.

\acknowledgments
We thank S. V. Buldyrev, C. Edling, P. Gopikrishnan, P. Hedstr\"om,
M. Macy, and V. Plerou for stimulating discussions and A. L. Stinchcombe 
for numerous comments and suggestions. FL thanks STINT (97/1837) and HSFR
(F0688/97). The CPS is supported by NSF and NIH.



\begin{thebibliography}{99}
\bibitem{Source} The primary data source was made available to us by
the Swedish Social Science Data Service at G\"oteborg. The data
collection was coordinated by Car-G\"oran Andr\ae \/ and Sven
Lundkvist.
\end{thebibliography}
\end{document}